\documentclass[robotics,article,accept,moreauthors,LaTeX]{Definitions/mdpi} 

\graphicspath{{Definitions/Plots/}}

\firstpage{1} 
\pubvolume{xx}
\issuenum{1}
\articlenumber{5}
\pubyear{2019}
\copyrightyear{2019}
\history{Received: 19 November 2019; Accepted: 29 December 2019; Published: date}
\updates{yes} 

\Title{Dedicated Nonlinear Control of Robot Manipulators in the Presence of External Vibration and Uncertain~Payload}

\Author{Mustafa M. Mustafa
~$^{1,}$*\orcidA{}, Ibrahim Hamarash $^{1,2}$ and Carl D. Crane
~$^{3}$}

\AuthorNames{Mustafa M. Mustafa, Ibrahim Hamarash and Carl D. Crane}

\address{%
$^{1}$ \quad Department of Electrical Engineering, Salahaddin University-Erbil, Erbil 44002, Kurdistan Region, Iraq\\
$^{2}$ \quad Department of Computer Science and Engineering, University of Kurdistan Hewler, Erbil 44001, Kurdistan Region, Iraq; ibrahim.hamad@ukh.edu.krd
 \\
$^{3}$ \quad Department of Mechanical and Aerospace Engineering, University of Florida, Gainesville, FL 32611, USA; ccrane@ufl.edu }

\corres{Correspondence: mustafa.atrushi@su.edu.krd; Tel.: +964-750-4968787}

\abstract{Robot manipulators are often tasked with working in environments with vibrations and are subject to load uncertainty. Providing an accurate tracking control design with implementable torque input for these robots is a complex topic. This paper presents two approaches to solve this problem. The approaches consider joint space tracking control design in the presence of nonlinear uncertain torques caused by external vibration and payload variation. The properties of the uncertain torques are used in both approaches. The first approach is based on the boundedness property, while the second approach considers the differentiability and boundedness together. The controllers derived from each approach differ from the perspectives of accuracy, control effort, and disturbance properties. A Lyapunov-based analysis is utilized to guarantee the stability of the control design in each case. Simulation results validate the approaches and demonstrate the performance of the controllers. The derived controllers show stable results at the cost of the mentioned properties.}


\begin{document}

\section{Introduction}
\label{sec:introduction}
Many robot manipulators work in environments with vibrations and are concurrently subject to payload uncertainty. These two factors cause a significant nonlinear disturbance torque that affects the dynamic behavior of the controllers, which complicates the issue of tracking control~\cite{du2016markerless}. Therefore, considering the disturbance while controlling the position of manipulators has been a topic of interest in recent~years.

Some researchers count the disturbance as part of uncertain dynamics while dealing with trajectory tracking control. Others consider the dynamics as known and deal with the disturbance separately. For~this purpose, a~variety of linear and nonlinear control tools have been tested. In~\cite{slotine1987adaptive}, the~trajectory tracking problem was treated in the presence of an unknown payload using adaptive control theory. The~same theory was used in~\cite{craig1987adaptive} to suppress disturbances and track desired trajectories uniformly by assuming that an accurate dynamic model of the manipulator is available. The~authors of~\cite{hsia1991robust} considered the uncertain load as part of unmodeled dynamics and represented it by a single disturbance force to be rejected utilizing a proportional derivative (PD control law). Dawson~et~al.~\cite{dawson1990robust} depended on the definition that the disturbance can be bounded~\cite{craig1987adaptive} to show that if the PD controller gains are chosen to be greater than that bound plus the other uncertain dynamics and the initial value of the tracking error is considered, then that error is uniformly bounded (UB). 

In~\cite{spong1992robust}, the~author derived a control law and proved that the tracking error is uniformly ultimately bounded (UUB) depending on the inertia parameters of the robot. The~control effort in~\cite{slotine1987adaptive, craig1987adaptive} is superior to that in~\cite{hsia1991robust, dawson1990robust, spong1992robust}, but~this kind of controller performs poorly in the presence of disturbances, such as unknown payload or external vibration. Some other works~\cite{slotine1985robust, yeung1988new} used the theory of variable structure (VSS) to drive the tracking error to a switching surface so that the disturbance will not affect the tracking control (i.e., asymptotic result). However, the~drawback of controllers that use VSS theory is the discontinuity/high-frequency of the control input, which makes it difficult to implement. Furthermore, other works take into account the case that the disturbance cannot be linearly parameterized and thus incorporate function approximation methods (neural network and fuzzy logic) into the above mentioned works and presented UUB results for the tracking~error. 

In~\cite{nafia2018robust}, the~fuzzy logic method was used to develop a robust tracking control for robot manipulators in the presence of unknown perturbations. In~\cite{leahy1991neural}, the~authors showed that a combination of robust model-based control and neural network payload estimation has the potential to provide payload-invariant high-speed trajectory tracking. In~\cite{kwan1998robust},  a~universal robust neural network controller was presented for motion control of rigid-link electrically-driven robots with unknown dynamics. This controller does not require an off-line training-phase, as~compared with other neural network approaches. Gao, H.~et~al.~\cite{gao2018neural} developed a neural network controller for vibration suppression of a two-link flexible robotic manipulator and they showed UUB results. The~mentioned works require velocity measurements to design the controllers. Further research has been conducted to eliminate the measurement of velocity and show global asymptotic and global UUB for joint position tracking~\cite{zhang2000global, dixon2004global}.
However, these controllers are restricted by many assumptions during the design~process. 

To overcome these restrictions, other efforts utilize the differentiable property of the uncertain terms and/or use robust integral of the sign of the error (RISE) feedback to develop new controllers for high-order dynamic systems and many other systems. In~\cite{cai2006robust}, a~robust adaptive control is presented for a multiple-input–multiple-output system that guarantees the tracking error is asymptotically driven to zero in the presence of bounded disturbances with bounded time derivatives. The~same technique was used in~\cite{patre2006asymptotic, makkar2007lyapunov, patre2008asymptotic} to reject the additive disturbance and the friction model that contains uncertain nonlinear parameterizable terms for Euler–Lagrange~systems. 

The RISE feedback was used in~\cite{shao2018rise, su2019rise} for trajectory tracking control and active vibration control for quadrotor unmanned aerial vehicles (UAVs) and flexible refueling hose, respectively. In~the same manner, the~RISE feedback was used in~\cite{pedroza2014robust} for suppressing limit cycle oscillations in UAVs with dynamic model uncertainty and parametric actuator uncertainty. Furthermore,  the~authors of~\cite{fischer2014nonlinear} used RISE feedback to compensate for uncertain, nonautonomous disturbances for a class of coupled, fully-actuated underwater vehicles. These controllers are continuous and yield asymptotic tracking errors.
Some other studies focus on disturbance torques that are caused by vibration and payload~variation. 

In~\cite{economou2000robust}, the~authors use finite impulse
response FIR 
 digital filters to design a controller that is robust to vibration and payload variation. However, digital filters work better for linear, time-invariant systems and they have some limitations with nonlinear systems. Additionally, this controller requires knowledge of the natural frequencies of the payload. In~\cite{mamani2012sliding}, a~sliding mode control was applied to solve the position tracking problem of a single-link flexible robot arm. The~controller is robust with regards to payload and actuator friction changes; however, it requires a high bandwidth actuator, which increases the control effort. V, Feliu.~et~al.~\cite{feliu2013robust} designed a control algorithm to cancel vibrations that are originated by the structural flexibility of the manipulator during movement. The~ control algorithm works efficiently for trajectory tracking without saturating the actuators, however, it requires a perfect kinematic model and a perfect knowledge of the compliance~matrix.

This paper studies the regulation problem in joint space for a manipulator subjected to uncertain torque due to vibration and payload variation. For~this objective, we consider that there is no uncertainty in the kinematics of the manipulator. Therefore, the~effect of the vibration and the payload variation are modeled as bounded disturbance torques~\cite{gu2005robust} to the joints, which are driving the links of the robot manipulator. Two approaches are used to design the control input depending on the characteristics of the disturbance torques based on Lyapunov analysis. The~first approach considers the vibration and payload variation torques are bounded. The~controller derived from this approach introduces UUB output for the tracking error with an acceptable control effort. The~second approach exploits the differentiable property plus the boundness of the disturbance torques. Therefore, the~controller derived from the second approach shows that the error goes to zero as time goes to infinity depending on the initial condition of the state (i.e., semiglobal stable result of the tracking error).
The performance of the controllers is evaluated as~follows:
\begin{enumerate}[leftmargin=*, labelsep=4.9mm]
    \item The controller derived from the first approach is compared with a PD controller in the presence of bounded disturbance torques that are caused by vibration and payload variation. The~simulation results show a UUB tracking error for this controller with a good control effort, while the PD controller performs poorly in term of~accuracy.
    
    \item The controller derived from the second approach is also compared with the PD controller in the presence of a differentiable and bounded vibration and payload variation torques and with a specific initial condition. The~simulation results show an asymptotic tracking error for this controller with low control effort. The~PD controller behaves almost the same as previously.
\end{enumerate}

In comparison with the mentioned existing advanced techniques, the~proposed approaches are conducted based on the best compact dynamic model of a rigid-link robot arm in the presence of disturbance torques due to vibration and payload variation. The~effects of both the vibration and payload variation are considered as uncertain disturbance torques. That is, in~the control implementation for both approaches, we do not require prior knowledge about the frequencies of the vibration and payload variation; however, high position tracking accuracy is achieved. Moreover, the~theoretical analysis presents important features in the performance of both controllers. Where the first approach gives less conservative tracking error. Simply put, it ensures the tracking error to be arbitrarily small through two constants that are injected into the controller. While the second approach introduces an arbitrarily high rate of convergence for the tracking error; however, the~convergence rate can be easily adjusted by tuning the gains in the control law.

The paper is organized as follows. In~Section~\ref{secdp}, the~dynamic model of the robot and some preliminaries are introduced. In~Section~\ref{secca}, the~problem formulation is defined, the~control approaches are presented, and~the controllers are designed. Simulations are carried out to illustrate the performance of the proposed controllers, and~the results are discussed in Section~\ref{secsr}. The~conclusions are given in~Section~\ref{secc}.
\section{Specified Dynamic Model and~Preliminaries}
\label{secdp}
Based on the control objective of this work, the~dynamics for an $n$-degree-of-freedom, revolute-joints, serial robot arm in the presence of external vibration and uncertain payload can be expressed in joint-space coordinates as follows:
\begin{equation}
    M(q)\ddot{q}+V_{m}(q,\dot{q})\dot{q}+G(q)=\tau-\tau_{v}-\tau_{l} \label{eq:1}
\end{equation}
where $q$, $\dot{q}$, and~$\ddot{q}$ $\in$ $\Re^{n}$ denote the angular displacement, angular velocity, and~angular acceleration vectors, respectively;
$M(q)$ $\in$ $\Re^{n\times n}$ represents the inertia matrix; $V_{m}(q,\dot{q})$ $\in$ $\Re^{n\times n}$ represents
the Coriolis/centripetal matrix; $G(q)$ $\in$ $\Re^{n}$ represents the gravity vector; $\tau$ $\in$ $\Re^{n}$ represents the torque input of the joints; $\tau_{v}$ $\in$ $\Re^{n}$ represents the disturbance due to external vibration; and $\tau_{l}$ $\in$ $\Re^{n}$ represents the load~torque.

For the purpose of controller design and analysis, we state some properties and assumptions that are used in this work. The~properties are applicable for the dynamics of a standard fixed-base n-link, revolute, direct-drive robot manipulator in Equation 
~\eqref{eq:1} \cite{from2010boundedness, dixon2013nonlinear}:

\begin{Property}
The  inertia matrix $M(q)$ is a positive-definite and symmetric matrix that satisfies
\begin{equation}
   \lambda_m \left\Vert \chi \right\Vert ^2 \leq \chi^T M(q) \chi \leq  \lambda_M \left\Vert \chi \right\Vert ^{2}, \; \forall\; \chi \in \Re^n 
   \label{eq:2}
\end{equation}
where  $\chi$ is a vector; $\lambda_m$ and $\lambda_M$ $\in \Re$ are known, positive, real, and~bounding constants; and $\left\Vert \cdot \right\Vert$ is the Euclidean norm.
\end{Property}

\begin{Property}
The inertia and  Coriolis/centripetal matrices satisfy the following skew-symmetric relationship:
\begin{equation}
\chi^T(\dot{M}(q)-2V_{m}(q,\dot{q}))\chi=0, \;\forall\; \chi \in \Re^n
\label{eq:3}
\end{equation}
where  $\dot{M}(q)$ is the derivative of the inertia matrix.
\end{Property}

\begin{Assumption}
The friction effects are beyond the scope of this study; hence these effects are neglected in \eqref{eq:1}.
\end{Assumption}
\section{Problem Formulation and Control~Approaches}
\label{secca}
\subsection{Problem~Formulation}
We expect that a robot manipulator can perform its task even in the presence of an external vibration and/or a payload variation. Hence, we need to regulate the robot's end-effector trajectory in order to follow a desired trajectory in the work-space. The~task-space trajectory of the end-effector can be determined from the joint-space trajectory by applying the kinematic analysis~\cite{crane2008kinematic}. Therefore, the~objective is to design the joint's torque input $\tau$ for the dynamic model in Equation \eqref{eq:1} that can make the angular displacement $q$ follow the desired trajectory $q_d$ (i.e., $q_d-q \to 0$ as $t \to \infty$, where $t$ denotes the time) from the perspective that there are disturbance torques of external vibration $\tau_v$ and payload variation $\tau_l$.

To expedite the analysis that leads to the design, we introduce the following terms~\cite{slotine1991applied}:
\begin{equation}
\eta_1:=\varDelta \dot{q} + \sigma_1\varDelta q
\label{eq:5}
\end{equation}
\begin{equation}
\eta_2:=\dot{\eta_1} + \sigma_2\eta_1
\label{eq:6}
\end{equation}
where $\eta_1$ and $\eta_2$ $\in \Re^n$ are filtered signals of $\varDelta q$ and $\eta_1$, respectively; $\varDelta q:=q_d-q$ is the difference between the desired and current angular displacement of the joint; $\varDelta \dot{q}$ is the derivative of $\varDelta q$; $\sigma_1$ and $\sigma_2$ $\in \Re^{n\times n}$ are positive-definite diagonal gain~matrices.

\begin{Assumption}
The desired angular displacement of the joint $q_d$ is known; $q$ and $\dot q$ are measurable, which makes $\eta_1$, $\dot{\varDelta q}$, and~$\varDelta q$ measurable as well.
\end{Assumption}

\begin{Assumption}
The angular acceleration $\ddot{q}$ is not measurable, which makes $\eta_2$ not measurable, since $\ddot{q}$ is a dependent term in \eqref{eq:6}.
\end{Assumption}
\subsection{Control Approach Based on the Bounded-Disturbance: First Control~Approach}
\label{sec:ap1}
In order to design a controller that fulfills our objective, we use the Lyapunov function candidate in Equation \eqref{eq:7}
\begin{equation}
    V_1:=\frac{1}{2}\eta_1^T M(q)\eta_1
    \label{eq:7}
\end{equation}
where $V_1$ is a scalar (i.e., $V_1 \in \Re$). The~time derivative of \eqref{eq:7} gives
\begin{equation}
\begin{split}
\dot {V_1} & =\frac{1}{2}\dot{\eta_1}^T M(q)\eta_1+\frac{1}{2}\eta_1^T\dot{M}(q)\eta_1+\frac{1}{2}\eta_1^T M(q)\dot{\eta_1}=\frac{1}{2}\eta_1^T\dot{M}(q)\eta_1+\eta_1^T M(q)\dot{\eta_1}.
    \label{eq:8}
    \end{split}
\end{equation}

Motivated by the term $M(q)\dot{\eta_1}$ in \eqref{eq:8}, the~following equation can be obtained by taking the derivative of \eqref{eq:5} and then multiplying it by $M(q)$:
\begin{equation}
M(q)\dot{\eta_1}=M(q)\varDelta\ddot q+\sigma_1 M(q) \varDelta \dot{q}.
    \label{eq:9}
\end{equation}

Equation \eqref{eq:9} can be extended by substituting $\varDelta\ddot q=\ddot{q}_d-\ddot{q}$ to obtain
\begin{equation}
M(q)\dot{\eta_1}=M(q)\ddot{q}_d-M(q)\ddot{q}+\sigma_1 M(q) \varDelta \dot{q}.
    \label{eq:10}
\end{equation}

Using $M(q)\ddot{q}=\tau-\tau_{v}-\tau_{l}-V_{m}(q,\dot{q})\dot{q}-G(q)$ from \eqref{eq:1}, $\dot{q}=\dot{q}_d-\varDelta\dot q$, and~$\varDelta\dot q=\eta_1-\sigma_1 \varDelta q$, we can write \eqref{eq:10} as

\begin{equation*}
\begin{split}
M(q)\dot{\eta_1} = &M(q)\ddot{q}_d-\tau+\tau_{v}+\tau_{l}+V_{m}(q,\dot{q})\dot{q} +G(q)+\sigma_1 M(q) \varDelta \dot{q}.
    \end{split}
\end{equation*}

Then
\begin{equation}
\begin{split}
M(q)\dot{\eta_1} = & M(q)\ddot{q}_d+V_{m}(q,\dot{q})[\dot q_d-\eta_1+\sigma_1 \varDelta q] +G(q)+\sigma_1 M(q) \varDelta \dot{q}-\tau+\tau_{v}+\tau_{l}.
    \label{eq:11}
    \end{split}
\end{equation}

\begin{Assumption}
The external vibration and payload variation torques, and~the gravity term in \eqref{eq:1} are upper bounded as
\begin{equation}
\left\Vert \tau_{v}+\tau_{l} \right\Vert\leq v_{b}+l_{b},\;\;\; \left\Vert G(q) \right\Vert\leq g_b
\label{eq:4}
\end{equation}
where $v_{b}$, $l_{b}$, and~$g_b$ indicate known, positive, and~real constants.
\end{Assumption}

Depending on the above assumption, the~context of Lyapunov stability~\cite{khalil2002nonlinear}, and~the fact that it is possible to include the measurable and/or known terms in the input torque expression, we design $\tau$ as
\begin{equation}
    \begin{split}
    \tau  = &K_1\eta_1+\frac{(v_b+l_b)^2}{d}\eta_1+M(q)\ddot{q}_d+V_{m}(q,\dot{q})\dot{q}_d+V_{m}(q,\dot{q})\sigma_1 \varDelta q +G(q)+\sigma_1 M(q) \varDelta \dot{q}
    \label{eq:12}
    \end{split}
\end{equation}
where $K_1=K_1^T>0$ denotes the control gain diagonal matrix; $d$ denotes a design parameter; $v_b$ and $l_b$ are introduced in Assumption 4. The~term $\frac{(v_b+l_b)^2}{d}\eta_1 \in \Re^n$ in \eqref{eq:12} is considered as an auxiliary control~signal.

The validity of the designed $\tau$ in \eqref{eq:12} can be investigated by substituting \eqref{eq:12} into \eqref{eq:11} and then~\eqref{eq:11} into \eqref{eq:8}, which gives
\begin{equation}
    \begin{split}
    \dot V_1 = & \frac{1}{2}\eta_1^T\dot{M}(q)\eta_1-\eta_1^T V_{m}(q,\dot{q})\eta_1 +\eta_1^T[\tau_v+\tau_l-\frac{(v_b+l_b)^2}{d}\eta_1-K_1\eta_1].
    \label{eq:13}
    \end{split}
\end{equation}

\textls[-20]{Applying the skew-symmetric relationship in Property 2 and the inequalities in Assumption 4, \eqref{eq:13} can be written as}
\begin{equation}
\dot V_1 \leq{(v_b+l_b)} \left\Vert{\eta_1}\right\Vert-\frac{(v_b+l_b)^2}{d}{\left\Vert{\eta_1}\right\Vert}^2-{\eta_1^T K_1 \eta_1}.
\label{eq:14}
\end{equation}

Then, \eqref{eq:14} yields
\begin{equation}
\dot V_1 \leq-{\eta_1^T K_1 \eta_1}+(v_b+l_b) \left\Vert{\eta_1}\right\Vert{[1-\frac{(v_b+l_b)}{d}{\left\Vert{\eta_1}\right\Vert}]}.
\label{eq:15}
\end{equation}

\begin{Assumption}
Let $\mu:=d/(v_b+l_b)$; the norm value of the filtered signal $\eta_1$ is considered to be
\begin{equation}
\left\Vert{\eta_1}\right\Vert > \mu > 0.
\label{eq:16}
\end{equation}
\end{Assumption}

In practice, the~inequality in \eqref{eq:16} is ascertained by initializing $\eta_1$ in \eqref{eq:5} suitably. This can be obtained through the initial value of the error. Simply put, since the desired joint displacement $q_d$ is known and the actual joint displacement $q$ is measurable, the~error $\Delta q$ can be easily initiated to be greater than zero (i.e., $\Delta q(0)>0$). Consequently, the~positive-definite diagonal gain matrix $\sigma_1$ guarantees $\eta_1(0)>0$ based on the definition of $\eta_1$ in \eqref{eq:5}; hence $\left\Vert{\eta_1}\right\Vert >0$ according to the properties of the Euclidean norm~\cite{desoer1975feedback}. Therefore, when the design parameter $d$ is selected to be arbitrarily small compared to $(v_b+l_b)$, the~inequality in \eqref{eq:16} will be satisfied and provides sufficient conditions for the stability. Thus, Equation \eqref{eq:15} can be simplified to
\begin{equation}
\dot V_1 \leq-{\eta_1^T K_1 \eta_1}.
\label{eq:17}
\end{equation}

The Lyapunov-like theorem of uniform and ultimate boundedness~\cite{khalil2002nonlinear} can be applied to upper bound $\left\Vert{\eta_1(t)}\right\Vert$. Using Property 1, the~Lyapunov function in \eqref{eq:7} can be upper and lower bounded by positive definite functions as follows
\begin{equation}
\alpha_1(\eta_1)  \leq V_1 \leq \alpha_2(\eta_1)
    \label{eq:18}
\end{equation}
where $\alpha_2(\eta_1)=(1/2)\lambda_m\left\Vert{\eta_1(t)}\right\Vert^2$, $\alpha_2(\eta_1)=(1/2)\lambda_M\left\Vert{\eta_1(t)}\right\Vert^2$; $\lambda_m$ and $\lambda_M$ are defined in \eqref{eq:2}.
The~term $\eta_1^T K_1 \eta_1$ in \eqref{eq:17} is a continuous positive definite function, $\forall \left\Vert{\eta_1(t)}\right\Vert \geq \mu > 0$, where $\mu$ is defined in~\eqref{eq:16}. Thus, according to the Lyapunov-like theorem,
\begin{equation}
\begin{split}
    \left\Vert{\eta_1(t)}\right\Vert & \leq \alpha_1^{-1} (\alpha_2(\mu)) \\
    & \leq \sqrt{\frac{\lambda_M}{\lambda_m}(\frac{d}{v_b+l_b})^2} \\
    & \leq \sqrt{\frac{\lambda_M}{\lambda_m}}(\frac{d}{v_b+l_b}),\; \forall\; t \geq 0.
    \label{eq:19}
    \end{split}
\end{equation}

From \eqref{eq:19}, we conclude that $\eta_1$ converges to a small neighborhood of zero according to the value of $d$. Furthermore, based on \eqref{eq:5}, $\varDelta q$ is simply a low-pass filter of $\eta_1$ signal~\cite{lewis2003robot}. Therefore, $\varDelta q$ is UUB with the ultimate bound result shown in \eqref{eq:19}. In~this way, the~practical deliverable input torque in \eqref{eq:12} yields an accurate tracking control for the robot joints.
\subsection{Control Approach Based on the Bounded-Differentiable-Disturbance: Second Control~Approach}
\label{sec:ap2}
As shown above, the~control approach drives the tracking error towards a small bound. Therefore, another approach will be followed in order to obtain an asymptotic result for the tracking~error.

A new Lyapunov function candidate is selected
\begin{equation}
    V_2:=\frac{1}{2}\eta_2^T M(q)\eta_2+\frac{1}{2}\eta_1^T
    \eta_1+\frac{1}{2}\varDelta {q}^T\varDelta q+W
    \label{eq:20}
\end{equation}
where $W \in \Re^+$ is an auxiliary function that will be formulated subsequently. The~derivative of~\eqref{eq:20}~gives
\begin{equation}
\dot{V_2}  = \frac{1}{2}\eta_2^T\dot{M}(q)\eta_1+\eta_2^T M(q)\dot{\eta_2}+\eta_1^T\dot{\eta_1}+\varDelta {q}^T\varDelta q
+\dot{W}.
    \label{eq:21}
\end{equation}

Analogous to the previous approach, \eqref{eq:6} is multiplied by $M(q)$ to obtain
\begin{equation}
\begin{split}
M(q)\eta_2 & = M(q)\dot{\eta_1} + M(q)\sigma_2\eta_1 \\
& = M(q)(\varDelta \ddot{q} + \sigma_1\varDelta \dot{q}) + \sigma_2 M(q) \eta_1 \\
& =  M(q) (\ddot{q}_d-\ddot{q}+\sigma_1\varDelta \dot{q})+\sigma_2 M(q) \eta_1.
 \label{eq:22}
\end{split}
\end{equation}

By substituting $M(q)\ddot{q}$ from \eqref{eq:1} into \eqref{eq:22}, $M(q)\eta_2$ can be expressed as
\begin{equation}
\begin{split}
M(q)\eta_2  = & M(q)\ddot{q}_d+V_m(q, \dot{q})\dot{q}+G(q) +\tau_v+\tau_l -\tau+ M(q) (\sigma_1\varDelta \dot{q}+\sigma_2 \eta_1).
 \label{eq:23}
\end{split}
\end{equation}

To facilitate the analysis process, the~terms in \eqref{eq:23} that are linear in the parameters can be separated and written as~\cite{craig1988adaptive}
\begin{equation}
Y(q,\dot{q},\ddot{q})\phi =  M(q)\ddot{q}_d+V_m(q,\dot{q})\dot{q}+G(q)
 \label{eq:24}
\end{equation}
where $Y(q,\dot{q},\ddot{q}) \in \Re^{n\times m}$ represents the matrix that is a function of $q$, $\dot{q}$ and $\ddot{q}$ vectorsand $\phi \in \Re^m$ represents the vector of constant parameters. The~matrix $Y$ can be expressed as a function of $q_d$, $\dot{q}_d$, and~$\ddot{q}_d$, which are considered to exist and be bounded. Therefore, \eqref{eq:24} can be expressed as
\begin{equation}
\begin{split}
{Y_d}(q_d,\dot{q}_d,\ddot{q}_d)\phi = &  M(q_d)\ddot{q}_d+V_m(q_d, \dot{q}_d)\dot{q}_d+G(q_d).
 \label{eq:25}
\end{split}
\end{equation}

By adding and subtracting \eqref{eq:25} to \eqref{eq:23}, we can rewrite \eqref{eq:23} as follows:
\begin{equation}
\begin{split}
M(q)\eta_2  = & (Y-Y_d)\phi + M(q) (\sigma_1\varDelta \dot{q}+\sigma_2 \eta_1) + {Y_d}\phi+ \tau_v+\tau_l-\tau.
 \label{eq:26}
\end{split}
\end{equation}

Now, we define a new auxiliary function as follows:
\begin{equation}
    Q := (Y-Y_d)\phi + M(q) (\sigma_1\varDelta \dot{q}+\sigma_2 \eta_1)
    \label{eq:27}
\end{equation}
where $Q(q,\dot{q},q_d,\dot{q}_d,\ddot{q}_d) \in \Re^n$ is a function of terms that can be upper~bounded.

\begin{Assumption}
The disturbance torques due to the exogenous vibration and payload variation are continuous, bounded, and~differentiable.
\end{Assumption}

Substituting \eqref{eq:27} into \eqref{eq:26} and then differentiating gives
\begin{equation}
M(q)\dot{\eta_2} + \dot{M(q)}\eta_2 = \dot{Q} + \dot{{Y_d}}\phi+ \dot{\tau_v}+\dot{\tau_l}-\dot{\tau}.
 \label{eq:28}
\end{equation}

Based on the analysis thus far and motivated by the idea of proportional derivative control, the~derivative of the torque input $\dot{\tau}$ is designed as follows:
\begin{equation}
\begin{split}
\dot{\tau}=K_2 \dot{\eta_1}+\dot{\eta_1}+K_2 \eta_2+\eta_2+K_3 sgn(\eta_1)
 \label{eq:29}
\end{split}
\end{equation}
where $K_2$ and $K_3 \in \Re$ are positive gains. Since we need the torque input $\tau$ in the dynamic system, $\dot{\tau}$~in~\eqref{eq:29} is designed to be integrable.
Substituting \eqref{eq:6} into \eqref{eq:29} and integrating both sides gives
\begin{equation}
\begin{split}
\tau =  & 2 K_2 \eta_1(t)+2\eta_1(t) +  \int_{0}^{t} [K_2 \sigma_2\eta_1(\epsilon)+\sigma_2\eta_1(\epsilon) +K_3 sgn(\eta_1(\epsilon))] d\epsilon.
 \label{eq:30}
\end{split}
\end{equation}

We know from \eqref{eq:5} that $\eta_1$ contains the joint displacement and velocity, which are measurable in this study, and~thus, the~term $\int_{0}^{t} [K_2\sigma_2\eta_1(\epsilon)+\sigma_2\eta_1(\epsilon) +K_3 sgn(\eta_1(\epsilon))] d\epsilon$ can be determined~numerically.

To investigate the stability of the system with the designed torque, we substitute \eqref{eq:29} into \eqref{eq:28} and apply some algebraic manipulations, which gives
\begin{equation}
\begin{split}
M(q)\dot{\eta_2} = & -\frac{1}{2} \dot{M}(q) \eta_2-\frac{1}{2} \dot{M}(q) \eta_2 + \dot{Q}-(K_2+1)\dot{\eta_1}+\eta_1+\dot{\tau_v}+\dot{\tau_l}\\
&+\dot{Y}_{d}\phi-(K_2+1)\eta_2-K_3 sgn(\eta_1)-\eta_1
 \label{eq:31}
\end{split}
\end{equation}
where $\eta_1$ is added and subtracted to facilitate the subsequent analysis. Using \eqref{eq:31} and \eqref{eq:5} and performing some simplifications, \eqref{eq:21} can be re-expressed as
\begin{equation}
\begin{split}
\dot{V} = &\eta_2^{T}[ -\frac{1}{2}\dot{M}(q)\eta_2+\dot{Q}-(K_2+1)\dot{\eta_1}+\eta_1 +\dot{\tau_v}+\dot{\tau_l}+\dot{Y}_{d}\phi-(K_2 + 1)\eta_2 - K_3 sgn(\eta_1)] \\
& +\varDelta q^T \eta_1-\sigma_1 \varDelta q^T \varDelta q - \sigma_2 \eta_1^{T} \eta_1 + \dot{W}.
 \label{eq:32}
\end{split}
\end{equation}

Based on Barbalat's lemma~\cite{khalil2002nonlinear}, the~condition that the auxiliary function $W$ should be positive and its derivative will be used to cancel some terms in \eqref{eq:32}. $W$ is formulated as follows:

\begin{equation*}
\begin{split}
W = & K_3 \left\Vert i_1 \right\Vert - i_1^T i_2  - \int_{0}^{t} \eta_2^T[\dot{\tau}_v (\epsilon)+\dot{\tau}_l (\epsilon)+\dot{Y}_{d} (\epsilon) \phi-K_3 sgn(\eta_1(\epsilon))]d\epsilon
\end{split}
\end{equation*}
where $i_1$ and $i_2 \in \Re^n$ are the values of $\eta_1$ and $(\dot{\tau_v}+\dot{\tau_l}+\dot{Y}\phi)$ at the initial conditions, respectively. In~order to guarantee that $W$ is always positive, the~gain $K_3$ should be selected to be greater than $(\dot{\tau_v}+\dot{\tau_l}+\dot{Y}\phi)$, which is possible since $\dot{\tau_v}$ and $\dot{\tau_l}$ are bounded according to Assumption 6, the~desired trajectories $q_d,\dot{q}_d,\ddot{q}_d$ are known, which makes $\dot{Y}_d$ known, and~$\phi$ is constant. Thus,
\begin{equation}
\dot{W}=-\eta_2^T(\dot{\tau_v}+\dot{\tau_l}+\dot{Y}_{d}\phi-K_3 sgn(\eta_1))
\label{eq:33}
\end{equation}
where all terms in $\dot{W}$ are uniformly continuous. Therefore, using \eqref{eq:33} $\dot{V}$ yields
\begin{equation}
\begin{split}
\dot{V} = & \eta_2^{T}[ -\frac{1}{2}\dot{M}(q)\eta_2+\dot{Q}-(K_2+1)\dot{\eta_1}+\eta_1] -\eta_2^T[(K_2+1)\eta_2+K_3 sgn(\eta_1)]  +\varDelta q^T \eta_1\\
& -\sigma_1 \varDelta q^T \varDelta q - \sigma_2 \eta_1^{T} \eta_1.
 \label{eq:34}
\end{split}
\end{equation}

We segregate and upper bound some terms in the above expressions based on~\cite{makkar2007lyapunov} as follows:

\begin{equation*}
\left\Vert -\frac{1}{2}\dot{M}(q)\eta_2+\dot{Q}-(K_2+1)\dot{\eta_1}+\eta_1\right\Vert
\leq \rho(\left\Vert z \right\Vert)\left\Vert z \right\Vert,
\end{equation*}
where $z \in \Re^{3n}$ is defined~as

\begin{equation*}
z = \left[\varDelta q^T \;\; \eta_1^T \;\; \eta_2^T\right]^T,
\end{equation*}
and by using the triangle inequality, we can obtain the following inequality:

\begin{equation*}
\varDelta q^T \eta_1 \leq \frac{1}{2}\left\Vert \varDelta q^T \right\Vert^2 + \frac{1}{2}\left\Vert \eta_1 \right\Vert^2.
\end{equation*}

We apply the above inequalities to upper bound $\dot{V}$ as follows:
\begin{equation}
\begin{split}
\dot{V} \leq & \left\Vert \eta_2 \right\Vert \rho(\left\Vert z \right\Vert)\left\Vert z \right\Vert - K_2\left\Vert \eta_2 \right\Vert^2  - \left\Vert \eta_2 \right\Vert^2 + \frac{1}{2}\left\Vert \varDelta q \right\Vert^2 + \frac{1}{2}\left\Vert \eta_1 \right\Vert^2   - \sigma_1 \left\Vert \varDelta q \right\Vert^2 - \sigma_2 \left\Vert \eta_1 \right\Vert^2.
 \label{eq:35}
\end{split}
\end{equation}

By applying the nonlinear damping proof~\cite{kokotovic1992joy}, we can~obtain

\begin{equation*}
\left\Vert \eta_2 \right\Vert \rho(\left\Vert z \right\Vert)\left\Vert z \right\Vert - K_2\left\Vert \eta_2 \right\Vert^2 \leq \frac{\rho^{2}(\left\Vert z \right\Vert)\left\Vert z \right\Vert^2}{K_2}.
\end{equation*}

Therefore, we can simplify \eqref{eq:35} as follows:
\begin{equation}
\begin{split}
\dot{V} \leq & \frac{\rho^{2}(\left\Vert z \right\Vert)\left\Vert z \right\Vert^2}{K_2} - [min(1, \sigma_1-\frac{1}{2}, \sigma_2-\frac{1}{2})]\left\Vert z \right\Vert^2.
 \label{eq:36}
\end{split}
\end{equation}

To obtain a negative definite result from the above expression, $K_2$ should be selected to be larger than $\rho^2$. Since $\rho$ is a function of the states that are bounded, we can obtain a semi-global stability result as the best depending on the initial conditions of the states. As~a result we can invoke Lemma 2 in~\cite{xian2004continuous} to show that $\varDelta q \rightarrow 0$ as $t \rightarrow \infty$.

\section{Simulation~Results}
\label{secsr}
The presented analytical results are evaluated via simulation. The~goal of the simulation is to demonstrate the performance of the proposed control approaches for a scenario where a robot manipulator is subjected to exogenous disturbance torques, which are caused by vibration and payload variation. For~this purpose, two signals are generated and combined to represent the disturbance torques due to vibration and payload variation. These signals are formed for the proposed approaches to match the properties and assumptions for each approach. In~order to represent this scenario, a~two-link robot manipulator is considered with the physical parameters in Table~\ref{tab1}. 

\begin{figure}[H]
\centering
	\includegraphics[width=\textwidth]{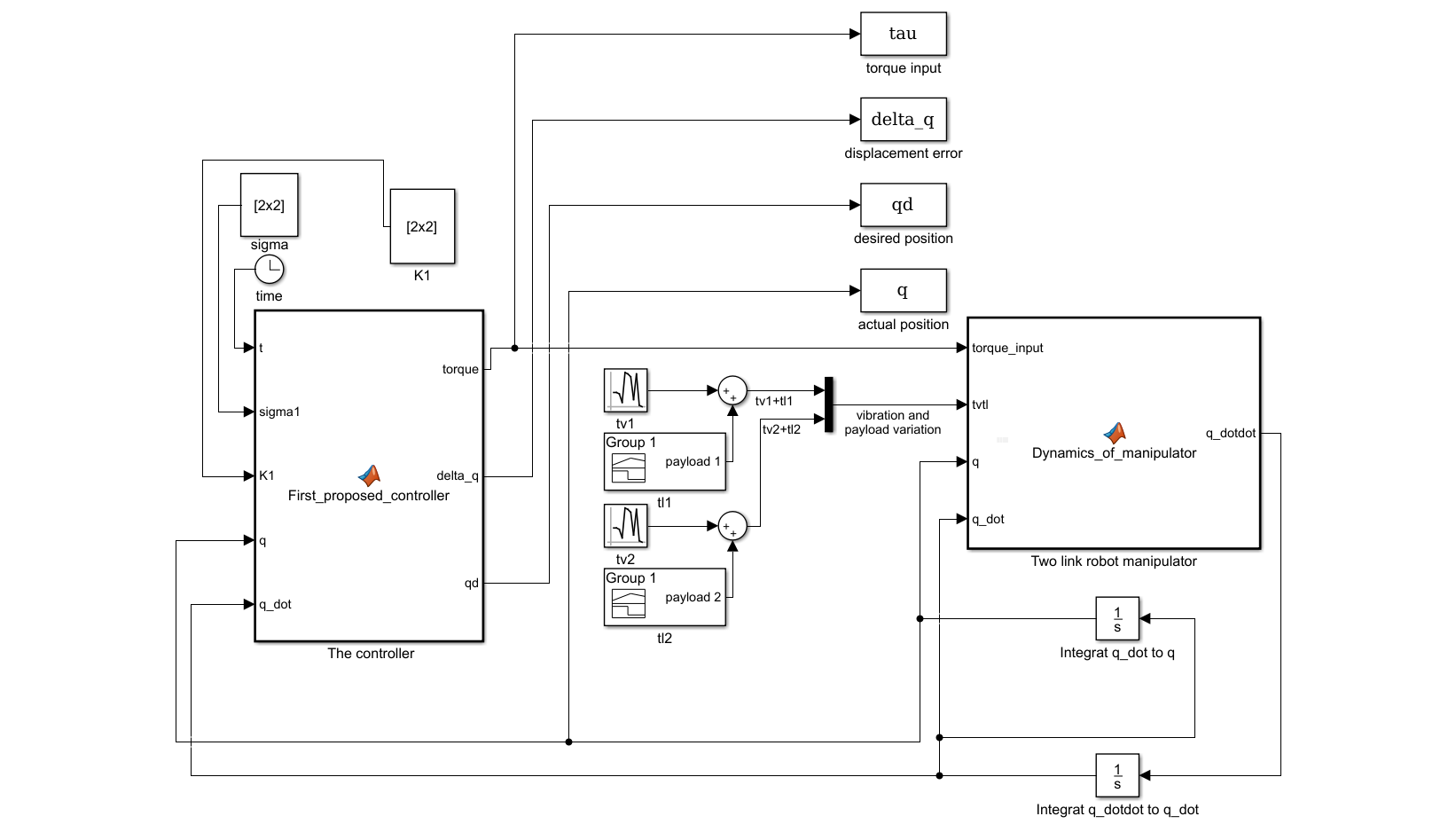}
\caption{Scheme of the simulated robot and the first~controller.\label{fig1}}
\end{figure}

\begin{table}[H]
\centering
\caption{Physical parameters of the two-link~manipulator.}
\setlength{\tabcolsep}{3pt}
\begin{tabular}{p{70pt}p{75pt}p{25pt}p{25pt}}
\toprule
\textbf{Symbol}& 
\textbf{Description
}&
\textbf{Value}&
\textbf{Unit} \\
\midrule
$m_1$& 
Mass of link 1& 
$0.5$& 
kg \\
$m_2$& 
Mass of link 2& 
$0.4$& 
kg \\
$l_1$& 
Length of link 1& 
$0.6$& 
m \\
$l_2$& 
Length of link 2& 
$0.5$& 
m \\
\bottomrule
\end{tabular}
\label{tab1}
\end{table}
 The two-link robot manipulator model, the~disturbance torques, and~the proposed controllers are simulated in MATLAB-SIMULINK as shown in Figure~\ref{fig1}. The~function that is used to integrate the states is ode45 (i.e., the~solver order of the integration is 45) and the sample time that is used to solve the integration is 0.53~ms.

The dynamics of the two-link robot manipulator can be expressed using \eqref{eq:1} as
\begin{equation}
\begin{split}
  \begin{bmatrix}
  M_{11} & M_{12} \\
  M_{21} & M_{22}
  \end{bmatrix}
  \begin{bmatrix}
  \ddot{q}_{1} \\
  \ddot{q}_{2}
  \end{bmatrix}
  + 
 \begin{bmatrix}
  V_{1} \\
  V_{2}
  \end{bmatrix} 
  +
\begin{bmatrix}
  G_{1} \\
  G_{2}
  \end{bmatrix}
    =
\begin{bmatrix}
  \tau_{1} \\
  \tau_{2}
  \end{bmatrix}
  -
  \begin{bmatrix}
  \tau_{v1} \\
  \tau_{v2}
  \end{bmatrix}
  - 
  \begin{bmatrix}
  \tau_{l1} \\
  \tau_{l2}
  \end{bmatrix}
  \label{eq:37}
  \end{split}
\end{equation}
where
\begin{equation*}
\begin{split}
    & M_{11}  =(m_1+m_2)l_{1}^2+m_2 l_{2}^2+2m_{2}l_1l_2cos(q_1)\\
    &
    M_{12}=m_2 l_{2}^2+m_2 l_1 l_2 cos(q_2)\\
    &
    M_{21}=m_2 l_{2}^2+m_2 l_1 l_2 cos(q_2)\\
    &
    M_{22}=m_2 l_{2}^2\\
    &
    V_1=-m_2l_1l_2(2\dot{q_1}\dot{q_2}+\dot{q_2}^2)sin(q_2)\\
    &
    V_2=m_2l_1l_2 \dot{q_1}^2 sin(q_2)\\
    &
    G_1=(m_1+m_2)gl_1cos(q_1)+m_2gl_2cos(q_1+q_2)\\
    &
    G_2=m_2gl_2cos(q_1+q_2)
    \end{split}
\end{equation*}
and $g=9.807$ m/s$^2$ is the Earth gravity constant. As~we will show subsequently, $\tau_v$ and $\tau_l$ are created for each of the proposed approaches according to their~assumptions.

The desired trajectories for both joints are assumed to be as follows:

\begin{equation*}
    \begin{bmatrix}
    q_{d1} \\
    q_{d2}
  \end{bmatrix}
  =  \begin{bmatrix}
    114.95^\circ \; sin(1.5t)\;e^{-0.03t}\\
    85.94^\circ \; cos(2t)\;e^{-0.03t}
  \end{bmatrix}
\end{equation*}
with the following initial conditions:
\begin{equation*}
    \begin{bmatrix}
    q_{d1}(0) \\
    q_{d2}(0)
  \end{bmatrix}
  =  \begin{bmatrix}
    0^\circ\\
    85.94^\circ
  \end{bmatrix}.
\end{equation*}

The proposed controllers are compared with a standard PD controller in order to demonstrate their performance over recently proposed controllers. The~input torque of the PD controller is designed~as
\begin{equation}
    \tau = K_p \varDelta q+K_d\varDelta \dot{q}
    \label{eq:38}
\end{equation}
where $K_p$ and $K_d \in \Re^{n \times n}$ denote the proportional and derivative diagonal gain matrices, respectively.

The proposed controllers are tested in terms of rejecting the uncertain disturbance torques due to vibration and payload variation. Beyond~the restriction on the disturbance properties, which are considered while formulating the disturbances, the~simulation shows that the purpose of each controller can be adequately achieved. The~control approaches in Sections~\ref{sec:ap1} and \ref{sec:ap2} are investigated in Sections~\ref{sec:1} and \ref{sec:2}, respectively. Moreover, quantitative analysis of the results is provided in Section~\ref{sec:3}.

\subsection{Simulation Results for the Control Approach Based on the Bounded~Disturbance}
\label{sec:1}
First, a~disturbance torque of vibration $\left[\tau_{v1} \;\;\tau_{v2}\right]^T$ by a \uppercase{G}aussian noise is composed taking into account that it should be bounded. Therefore, a~signal with a mean value $=\left[0 \;\; 0\right]^T$ Nm, a~variance $=\left[0.01 \;\; 0.015\right]^T$ Nm, and~a sampling time of $0.01$ s is set. Then, the~effect of the disturbance torque of the payload variation on the first and second joints is represented as follows:

\begin{equation*}
   \tau_{l1} = 
    \begin{cases}
0.65 \;\text{Nm,} & 4\;\text{s} \leq t \leq8\;\text{s} \\
0.15 \;\;\text{Nm,} & 8\;\text{s}\leq t \leq10\;\text{s}\\
0, \;\; & \text{otherwise}\\
\end{cases}
\end{equation*}

\begin{equation*}
   \tau_{l2} = 
    \begin{cases}
0.75 \;\text{Nm,} & 4\;\text{s} \leq t \leq8\;\text{s} \\
0.25 \;\;\text{Nm,} & 8\;\text{s}\leq t \leq10\;\text{s}\\
0, \;\; & \text{otherwise}.\\
\end{cases}
\end{equation*}

\begin{figure}[H]
\centering
	\includegraphics[scale=0.3]{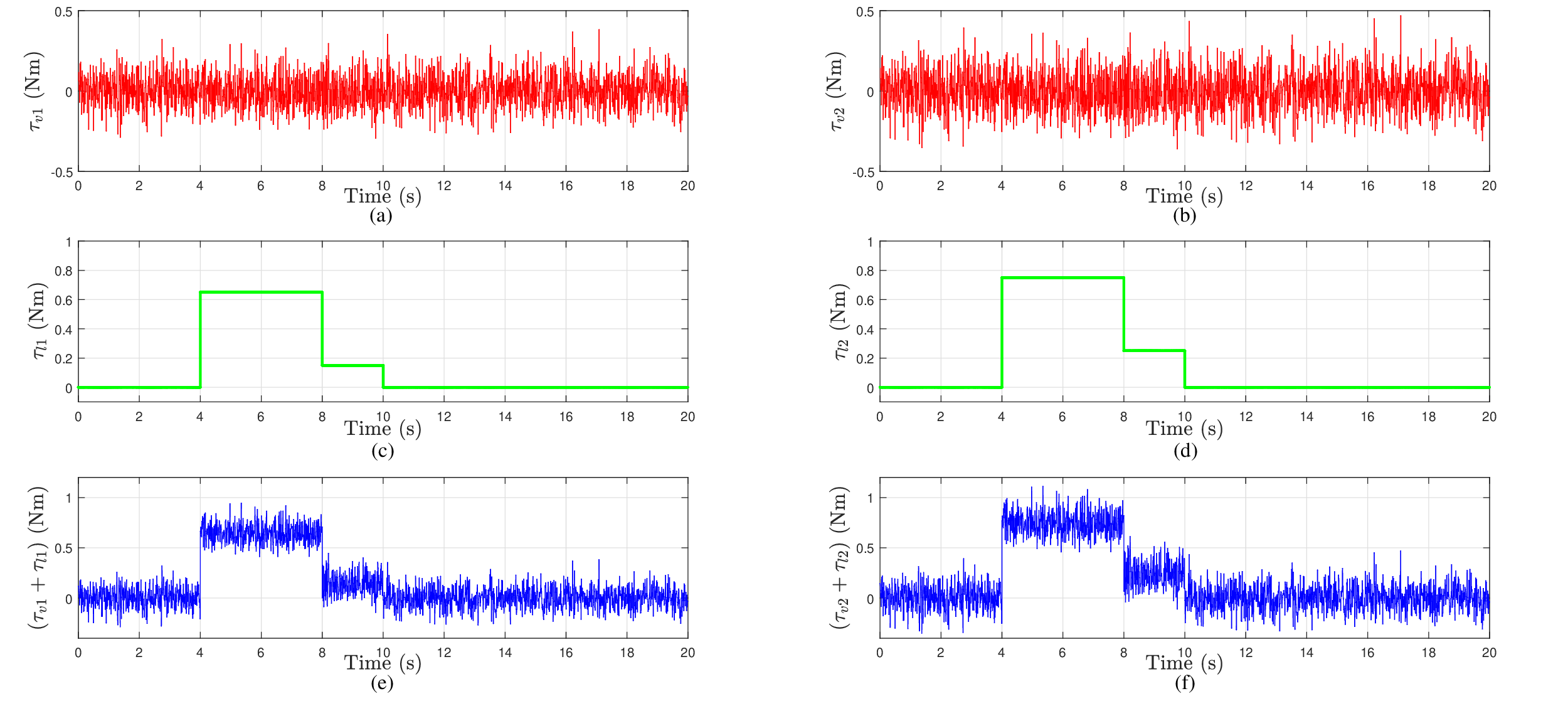}
\caption{\textls[-20]{Disturbance torques due to vibration and payload variation for the first approach: (\textbf{a},\textbf{b}) vibration effect; (\textbf{c},\textbf{d}) payload variation effect; and (\textbf{e},\textbf{f}) effect of combination on joint 1 and joint 2, respectively}.\label{fig2}}
\end{figure}

To provide further explanation, the~effects of both disturbances on the first and second joints are plotted in Figure~\ref{fig2}.
The combination of the disturbances is bounded as follows:

\begin{equation*}
    \begin{bmatrix}
    -0.2873\; \text{Nm} \\
    -0.3518\; \text{Nm}
  \end{bmatrix}
  \leq 
   \begin{bmatrix}
   \tau_{v1}+\tau_{l1}\\
    \tau_{v2}+\tau_{l2}
  \end{bmatrix}
    \leq
  \begin{bmatrix}
    0.9446\; \text{Nm} \\
    1.1108\;\text{Nm} 
  \end{bmatrix}.
\end{equation*}

\begin{figure}[H]
\centering
	\includegraphics[width=\textwidth]{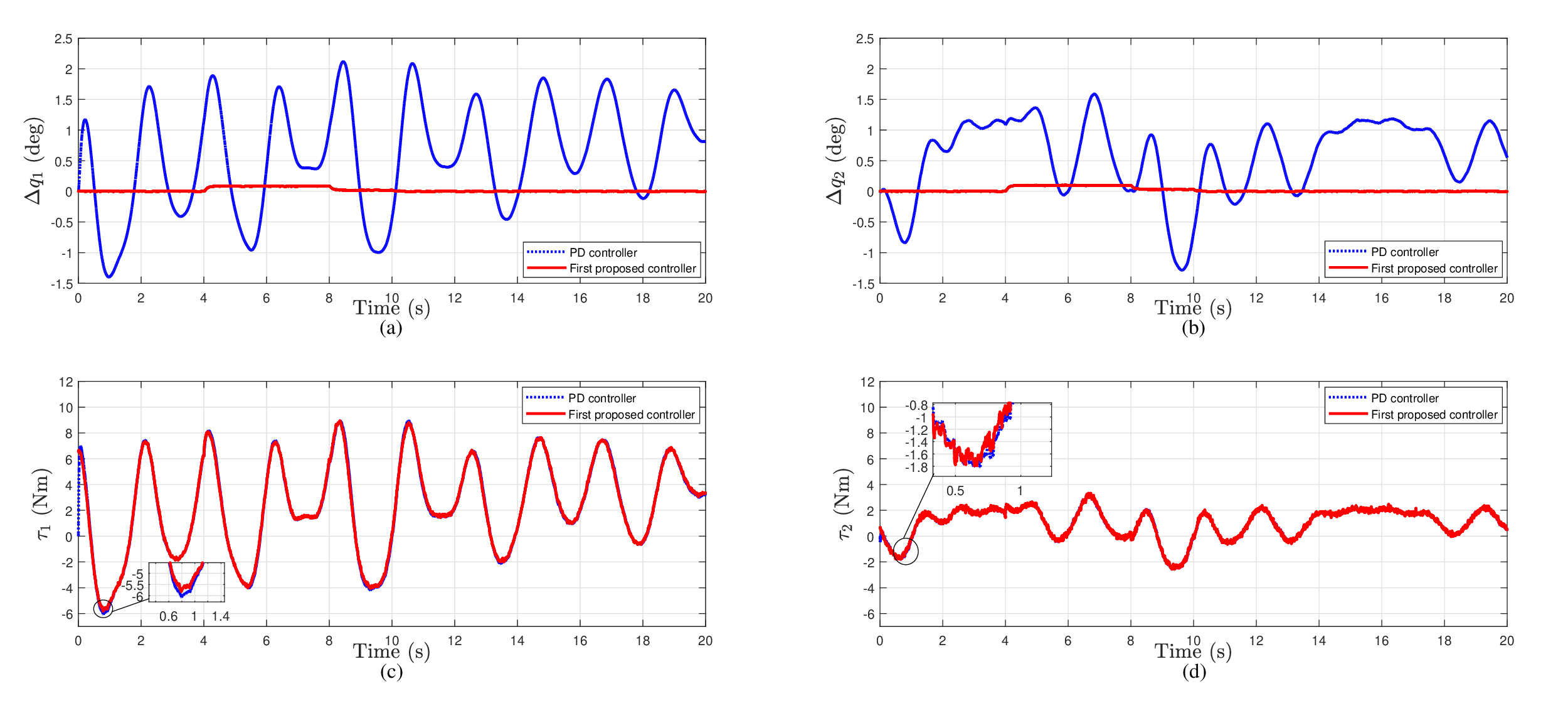}
\caption{{Tracking errors and control inputs for the first proposed approach compared with the proportional derivative (PD) approach: (\textbf{a},\textbf{b})~angular displacement errors of joint 1 and joint 2, respectively; (\textbf{c},\textbf{d}) input torques of joint 1 and joint 2, respectively.}\label{fig3}}
\end{figure}

\begin{figure}[H]
\centering
	\includegraphics[scale=0.35]{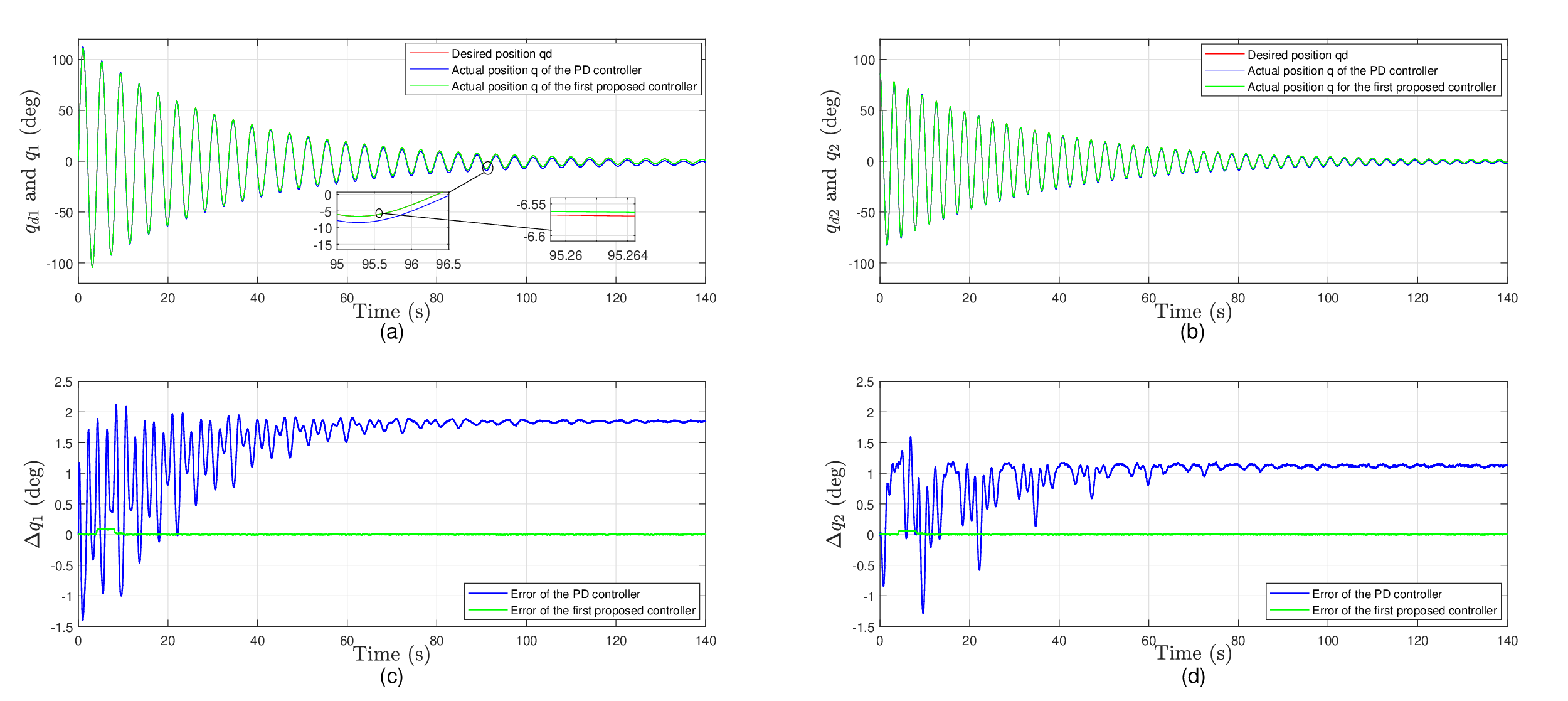}
\caption{Performance in terms of drifts and systematic errors for the first proposed approach compared with the PD approach:  (\textbf{a},\textbf{b}) desired and actual displacements of joint 1 and joint 2, respectively; (\textbf{c},\textbf{d})~angular displacement errors of joint 1 and joint 2, respectively.\label{fig4}}
\end{figure}

The control gains and design parameters in \eqref{eq:12} are selected as $K_1=$ diag$\left[40 \;\; 60\right]$, $\sigma_1=$ diag$\left[10 \;\; 12\right]$, $v_b+l_b=1.50$ Nm and $d=0.73$. The~proportional and derivative gains for the PD control input in \eqref{eq:38} are chosen to yield the best performance as $K_p=$ diag$\left[225 \;\; 100\right]$, $K_d=$ diag$\left[30 \;\; 20\right]$. The~tracking errors and input torques of this approach are shown in Figure~\ref{fig3}, where they are compared with the torques and errors obtained from the PD controller in \eqref{eq:38}. Although~the PD control can reject the disturbances, the~tracking errors are still large. By~contrast, the~proposed controller damps the same disturbances smoothly with less control efforts. Furthermore, the~proposed controller performs rather well in terms of drifts and systematic errors as well. To~show this, we run the simulation for 140 s and depict the actual displacements with their desired references and the errors between them in Figure~\ref{fig4}.
\unskip
\unskip
\subsection{Simulation Results for the Second Control Approach Based on the~Bounded-Differentiable-Disturbance}
\label{sec:2}
As in the previous part, the~disturbance signals are composed according to the assumptions of this approach. Therefore, we compose them~as

\begin{equation*}
   \begin{bmatrix}
   \tau_{v1}+\tau_{l1}\\
    \tau_{v2}+\tau_{l2}
  \end{bmatrix}
    =
  \begin{bmatrix}
   0.25\;sin(2t)+0.10\; \text{Nm} \\
   0.40\;sin(4t)+0.15\; \text{Nm} 
  \end{bmatrix}
\end{equation*}
where they are clearly bounded and differentiable. The~bounded values for the overall disturbance can be directly calculated from the above equation and~are 

\begin{equation*}
    \begin{bmatrix}
    -0.15 \;\text{Nm}  \\
    -0.25 \;\text{Nm} 
  \end{bmatrix}
  \leq 
   \begin{bmatrix}
   \tau_{v1}+\tau_{l1}\\
    \tau_{v2}+\tau_{l2}
  \end{bmatrix}
    \leq
  \begin{bmatrix}
    0.35\; \text{Nm} \\
    0.55\; \text{Nm} 
  \end{bmatrix}.
\end{equation*}

\begin{figure}[H]
\centering
	\includegraphics[scale=.32]{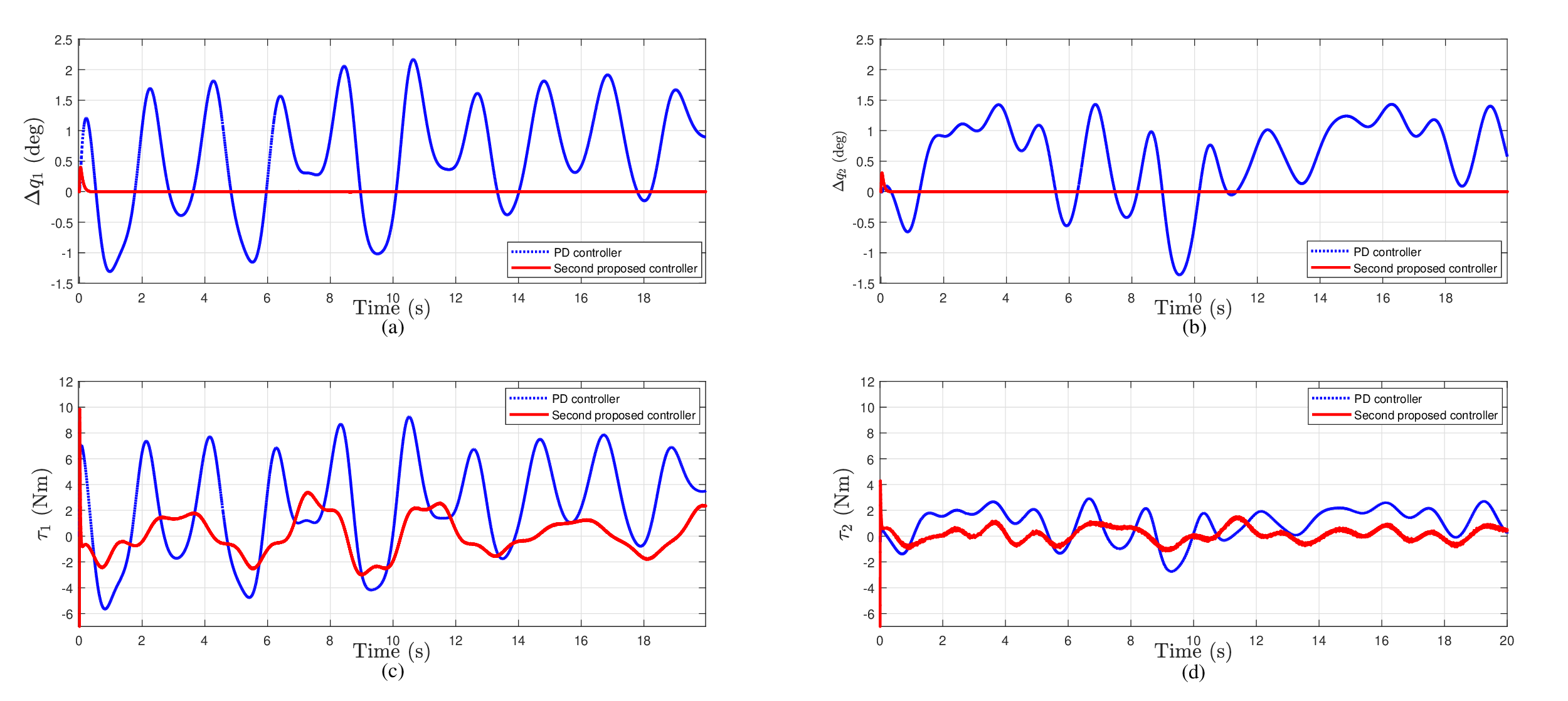}
\caption{Tracking errors and control inputs for the second proposed approach compared with the PD approach:  (\textbf{a},\textbf{b}) angular displacement errors of joint 1 and joint 2, respectively; (\textbf{c},\textbf{d}) input torques of joint 1 and joint 2, respectively.\label{fig5}}
\end{figure}

\begin{figure}[H]
\centering
	\includegraphics[width=\textwidth]{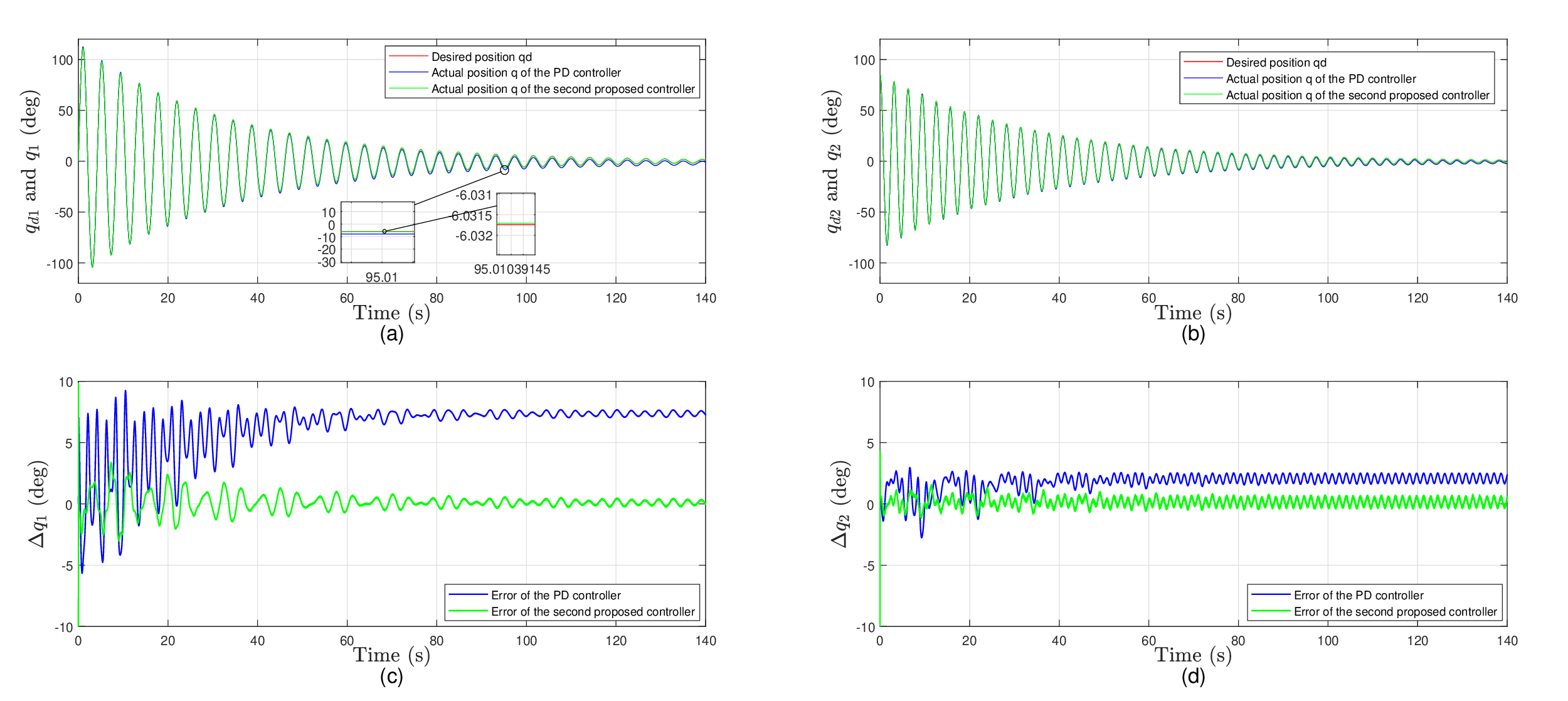}
\caption{Performance in terms of drifts and systematic errors for the second proposed approach compared with the PD approach:  (\textbf{a},\textbf{b}) desired and actual displacements of joint 1 and joint 2, respectively; (\textbf{c},\textbf{d}) angular displacement errors of joint 1 and joint 2, respectively.\label{fig6}}
\end{figure}

For this approach, we select the gains in \eqref{eq:30} as $K_2=75$, $K_3=100$, $\sigma_1=$ diag$\left[40 \;\; 50\right]$, and~$\sigma_2=$ diag$\left[30 \;\; 40\right]$. The~performance of the control inputs for this approach is compared with the PD control as shown in Figure~\ref{fig5}, where the gains and parameters of the PD control are the same as in Section~\ref{sec:1}.
This controller performs better in terms of accuracy and control effort because it drives the error to zero, but~it is restricted by Assumption 6. By~contrast, the~preceding controller drives the error to a value close to zero by only considering Assumption 1. Same as the preceding approach, the~performance of the proposed controller in terms of drifts and systematic errors is presented in~Figure~\ref{fig6}.

In both proposed approaches, the~torques have higher oscillations w.r.t. the PD controller; however, the~oscillations are not beyond the classical bandwidth limits of the actuators.
\subsection{Quantitative~Analysis}
\label{sec:3}
In order to further compare and validate the performance of the proposed controllers, the~following statistical indices are used:

\begin{enumerate}[leftmargin=*,labelsep=4.9mm]
    \item Maximum absolute value of the error for each joint.
\begin{equation}
     \varDelta {q_i}_{max} =  \max_{j=1,..,M} (\left| \varDelta {q_i}(j) \right|  ). 
     \label{eq:39}
    \end{equation}
    
    \item Root mean square (rms) values of the error and input torque for each joint.
\begin{equation}
     \varDelta {q_i}_{rms} =  \sqrt{\frac{1}{M} \sum_{j=1}^{M} \left\Vert \varDelta {q_i}(j) \right\Vert^2 }. 
     \label{eq:40}
    \end{equation}
\begin{equation}
     \tau_{i_{rms}} =  \sqrt{\frac{1}{M} \sum_{j=1}^{M} \left\Vert \tau_i(j) \right\Vert^2 }. 
     \label{eq:41}
    \end{equation}  
    
\item Percentage change in the rms values of the error and input torque for the proposed control approaches compared with the PD control.
\begin{equation}
     \%\;change \; \varDelta {q_i}_{rms} =  \frac{(\varDelta {q_i}_{rms})_{proposed \;controller}-(\varDelta {q_i}_{rms})_{PD\;controller}}{(\varDelta {q_i}_{rms})_{PD\;controller}}\times100\%, 
     \label{eq:42}
    \end{equation}
\begin{equation}
     \%\;change \;  \tau_{i_{rms}} =  \frac{( \tau_{i_{rms}})_{proposed \;controller}-( \tau_{i_{rms}})_{PD\;controller}}{( \tau_{i_{rms}})_{PD\;controller}}\times100\% 
     \label{eq:43}
    \end{equation}
    
\end{enumerate}
where $i$ is the joint number and $M$ is the number of sampling steps of the~simulation.

Tables~\ref{tab2} and \ref{tab3} summarize the performance for the both proposed approaches compared with the PD control. The~indices in
~\eqref{eq:39}--\eqref{eq:41} for both joints are obtained from the results in Figures~\ref{fig3} and~\ref{fig5}.
\begin{table}[H]
\centering
\caption{Performance summary for the first 
roposed approach compared with the PD~control.}
\setlength{\tabcolsep}{3pt}
\begin{tabular}{p{75pt}p{75pt}p{75pt}}
\toprule
\textbf{Indexes}& 
\textbf{PD Control}&
\textbf{First Approach}\\
\midrule
$\varDelta {q_1}_{max}$(deg)& 
$2.0981$&  
$0.0878$\\
$\varDelta {q_2}_{max}$(deg)& 
$1.5781$&  
$0.0590$\\
$\varDelta {q_1}_{rms}$(deg)& 
$1.0455$&  
$0.0374$\\
$\varDelta {q_2}_{rms}$(deg)& 
$0.8487$&  
$0.0253$\\
$\tau_{1_{rms}}$(Nm)& 
$4.3128$&  
$4.2933$\\
$\tau_{2_{rms}}$(Nm)& 
$1.5666$&  
$1.5719$\\
\bottomrule
\end{tabular}
\label{tab2}
\end{table}
\unskip

\begin{table}[H]
\centering
\caption{Performance summary for the second proposed approach compared with the PD~control.}
\setlength{\tabcolsep}{3pt}
\begin{tabular}{p{75pt}p{75pt}p{75pt}}
\toprule
\textbf{Indexes}& 
\textbf{PD Control}&
\textbf{Second Approach}\\
\midrule
$\varDelta {q_1}_{max}$(deg)& 
$2.1450$&  
$0.2198$\\
$\varDelta {q_2}_{max}$(deg)& 
$1.4325$&  
$0.1181$\\
$\varDelta {q_1}_{rms}$(deg)& 
$1.0459$&  
$0.0160$\\
$\varDelta {q_2}_{rms}$(deg)& 
$0.8636$&  
$0.0085$\\
$\tau_{1_{rms}}$(Nm)& 
$4.3161$&  
$1.4940$\\
$\tau_{2_{rms}}$(Nm)& 
$1.6011$&  
$0.5494$\\
\bottomrule
\end{tabular}
\label{tab3}
\end{table}

Table~\ref{tab4} presents the percentage changes in \eqref{eq:42} and \eqref{eq:43} for each one of the proposed controller compared with the PD controller.
Moreover, the~time for evaluating control actions of both controllers compared with the PD controller are presented in Table~\ref{tab5}.

\begin{table}[H]
\centering
\caption{Percentage changes of the errors and input torques for the proposed control approaches compared with the PD~control.}
\setlength{\tabcolsep}{3pt}
\begin{tabular}{p{75pt}p{75pt}p{75pt}}
\toprule
\textbf{Indexes}& 
\textbf{First Approach}&
\textbf{Second Approach}\\
\midrule
$\%\;change \; \varDelta {q_1}_{rms}$& 
$-96.40$&  
$-98.47$\\
$\%\;change \; \varDelta {q_2}_{rms}$& 
$-97.01$&  
$-99.01$\\
$\%\;change \;  \tau_{1_{rms}}$& 
$-0.45$&  
$-68.38$\\
$\%\;change \;  \tau_{2_{rms}}$& 
$+0.33$&  
$-96.565$\\
\bottomrule
\end{tabular}
\label{tab4}
\end{table}
\unskip
\begin{table}[H]
\centering
\caption{Performance in terms of computation time for the proposed control approaches compared with the PD~control.}
\setlength{\tabcolsep}{3pt}
\begin{tabular}{p{72pt}p{60pt}p{70pt}p{60pt}}
\toprule
\textbf{Controller}& 
\textbf{No. of Calls}&
\textbf{Time/Call (ms)}&
\textbf{Total Time (s)}\\
\midrule
PD controller& 
$20552$&  
$0.003$&
$0.062$\\
First controller& 
$76695$&  
$0.007$&
$0.536$\\
Second controller& 
$934670$&  
$0.005$&
$4.67$\\
\bottomrule
\end{tabular}
\label{tab5}
\end{table}
From the above tables, which compare the performance of each control approach with the PD control, we deduce the following~points:
\begin{enumerate}[leftmargin=*,labelsep=4.9mm]
    \item The first approach reduces the average tracking errors of both joints by about 97\% with almost the same control efforts.
    \item The second approach reduces the average tracking errors and control efforts of both joints by about 98\% and 81\%, respectively.
    \item The proposed controllers have higher computation time than that in the standard PD controller.
\end{enumerate}
\section{Conclusion}
\label{secc}
Two dedicated nonlinear approaches for tracking control of robot manipulators in joint space subject to uncertain torques due to vibration and payload variation are presented. In~both approaches, a~Lyapunov-based analysis is used to design the controllers and investigate their stability. The~first approach relieves the restrictions of the uncertain torques' characteristics and obtains a controller that yields a UUB tracking error. The~uncertain torques in this approach need only to be bounded. The~tracking error in the second approach yields an asymptotic result at the expense of an assumption that the uncertain torques are differentiable and bounded while considering the initial conditions of the states. Simulations are performed to validate the designed controllers and illustrate their performance in terms of accuracy, control efforts, and~computation time. The~designed controllers are compared with the PD controller and they show significant effect and superior control performance. The~first proposed controller reduces the average rms values of the tracking errors by around 97\% with almost the same rms values of the control efforts of the PD control. The~second proposed controller reduces the average rms values of the tracking errors and control efforts by around 98\% and 81\%, respectively. The~time necessary for evaluating control actions of the proposed controllers is higher than the time of the PD controller. However, this computation time is still less than that of advanced approaches and it is relatively small for modern CPUs.

\vspace{10pt}

\authorcontributions{Conceptualization, M.M.; formal analysis, M.M., C.C., and~I.H.; simulation code-writing, M.M.; supervision, C.C. and I.H.; writing-original draft, M.M. All authors have read and agreed to the published version of the manuscript.}


\funding{This research received no external~funding.}


\conflictsofinterest{The authors declare no conflict of~interest.} 

\reftitle{References}


\begin{thebibliography}{999}

\providecommand{\natexlab}[1]{#1}

\bibitem[Du \em{et~al.}(2016)Du, Zhang, and Liu]{du2016markerless}
Du, G.; Zhang, P.; Liu, X.
\newblock Markerless human--manipulator interface using leap motion with
  interval Kalman filter and improved particle filter.
\newblock {\em IEEE Trans. Ind. Inform.} {\bf 2016}, {\em
  12},~694--704.

\bibitem[Slotine and Li(1987)]{slotine1987adaptive}
Slotine, J.J.E.; Li, W.
\newblock On the adaptive control of robot manipulators.
\newblock {\em  Int. J. Robot. Res.} {\bf 1987}, {\em
  6},~49--59.

\bibitem[Craig \em{et~al.}(1987)Craig, Hsu, and Sastry]{craig1987adaptive}
Craig, J.J.; Hsu, P.; Sastry, S.S.
\newblock Adaptive control of mechanical manipulators.
\newblock {\em  Int. J. Robot. Res.} {\bf 1987}, {\em
  6},~16--28.

\bibitem[Hsia \em{et~al.}(1991)Hsia, Lasky, and Guo]{hsia1991robust}
Hsia, T.S.; Lasky, T.; Guo, Z.
\newblock Robust independent joint controller design for industrial robot
  manipulators.
\newblock {\em IEEE Trans. Ind. Electron.} {\bf 1991}, {\em
  38},~21--25.

\bibitem[Dawson \em{et~al.}(1990)Dawson, Qu, Lewis, and
  Dorsey]{dawson1990robust}
Dawson, D.; Qu, Z.; Lewis, F.; Dorsey, J.
\newblock Robust control for the tracking of robot motion.
\newblock {\em Int. J. Control} {\bf 1990}, {\em
  52},~581--595.

\bibitem[Spong(1992)]{spong1992robust}
Spong, M.W.
\newblock On the robust control of robot manipulators.
\newblock {\em IEEE Trans. Autom. Control} {\bf 1992}, {\em
  37},~1782--1786.

\bibitem[Slotine(1985)]{slotine1985robust}
Slotine, J.J.E.
\newblock The robust control of robot manipulators.
\newblock {\em  Int. J. Robot. Res.} {\bf 1985}, {\em
  4},~49--64.

\bibitem[Yeung and Chen(1988)]{yeung1988new}
Yeung, K.S.; Chen, Y.P.
\newblock A new controller design for manipulators using the theory of variable
  structure systems.
\newblock {\em IEEE Trans. Autom. Control} {\bf 1988}, {\em
  33},~200--206.

\bibitem[Nafia \em{et~al.}(2018)Nafia, El~Kari, Ayad, and
  Mjahed]{nafia2018robust}
Nafia, N.; El~Kari, A.; Ayad, H.; Mjahed, M.
\newblock Robust interval type-2 fuzzy sliding mode control design for robot
  manipulators.
\newblock {\em Robotics} {\bf 2018}, {\em 7},~40.

\bibitem[Leahy \em{et~al.}(1991)Leahy, Johnson, and Rogers]{leahy1991neural}
Leahy, M.; Johnson, M.A.; Rogers, S.K.
\newblock Neural network payload estimation for adaptive robot control.
\newblock {\em IEEE Trans. Neural Networks} {\bf 1991}, {\em
  2},~93--100.

\bibitem[Kwan \em{et~al.}(1998)Kwan, Lewis, and Dawson]{kwan1998robust}
Kwan, C.; Lewis, F.L.; Dawson, D.M.
\newblock Robust neural-network control of rigid-link electrically driven
  robots.
\newblock {\em IEEE Trans. Neural Netw.} {\bf 1998}, {\em
  9},~581--588.

\bibitem[Gao \em{et~al.}(2018)Gao, He, Zhou, and Sun]{gao2018neural}
Gao, H.; He, W.; Zhou, C.; Sun, C.
\newblock Neural network control of a two-link flexible robotic manipulator
  using assumed mode method.
\newblock {\em IEEE Trans. Ind. Informatics} {\bf 2018}, {\em
  15},~755--765.

\bibitem[Zhang \em{et~al.}(2000)Zhang, Dawson, de~Queiroz, and
  Dixon]{zhang2000global}
Zhang, F.; Dawson, D.M.; de~Queiroz, M.S.; Dixon, W.E.
\newblock Global adaptive output feedback tracking control of robot
  manipulators.
\newblock {\em IEEE Trans. Autom. Control} {\bf 2000}, {\em
  45},~1203--1208.

\bibitem[Dixon \em{et~al.}(2004)Dixon, Zergeroglu, and Dawson]{dixon2004global}
Dixon, W.E.; Zergeroglu, E.; Dawson, D.M.
\newblock Global robust output feedback tracking control of robot manipulators.
\newblock {\em Robotica} {\bf 2004}, {\em 22},~351--357.

\bibitem[Cai \em{et~al.}(2006)Cai, de~Queiroz, and Dawson]{cai2006robust}
Cai, Z.; de~Queiroz, M.S.; Dawson, D.M.
\newblock Robust adaptive asymptotic tracking of nonlinear systems with
  additive disturbance.
\newblock {\em IEEE Trans. Autom. Control} {\bf 2006}, {\em
  51},~524--529.

\bibitem[Patre \em{et~al.}(2006)Patre, MacKunis, Makkar, and
  Dixon]{patre2006asymptotic}
Patre, P.M.; MacKunis, W.; Makkar, C.; Dixon, W.E.
\newblock Asymptotic tracking for systems with structured and unstructured
  uncertainties.
\newblock  In Proceedings of the 45th I{EEE Conference on Decision and Control}, San Diego, CA, USA , December 2006 
;
  pp. 441--446.

\bibitem[Makkar \em{et~al.}(2007)Makkar, Hu, Sawyer, and
  Dixon]{makkar2007lyapunov}
Makkar, C.; Hu, G.; Sawyer, W.G.; Dixon, W.E.
\newblock Lyapunov-based tracking control in the presence of uncertain
  nonlinear parameterizable friction.
\newblock {\em IEEE Trans. Autom. Control} {\bf 2007}, {\em
  52},~1988--1994.

\bibitem[Patre \em{et~al.}(2008)Patre, MacKunis, Kaiser, and
  Dixon]{patre2008asymptotic}
Patre, P.M.; MacKunis, W.; Kaiser, K.; Dixon, W.E.
\newblock Asymptotic tracking for uncertain dynamic systems via a multilayer
  neural network feedforward and RISE feedback control structure.
\newblock {\em IEEE Trans. Autom. Control} {\bf 2008}, {\em
  53},~2180--2185.

\bibitem[Shao \em{et~al.}(2018)Shao, Meng, Liu, and Wang]{shao2018rise}
Shao, X.; Meng, Q.; Liu, J.; Wang, H.
\newblock RISE and disturbance compensation based trajectory tracking control
  for a quadrotor UAV without velocity measurements.
\newblock {\em Aerosp. Sci. Technol.} {\bf 2018}, {\em
  74},~145--159.

\bibitem[Su \em{et~al.}(2019)Su, Xie, and Li]{su2019rise}
Su, Z.; Xie, M.; Li, C.
\newblock RISE based active vibration control for the flexible refueling hose.
\newblock {\em Aerosp. Sci. Technol.} {\bf 2019}.

\bibitem[Pedroza \em{et~al.}(2014)Pedroza, MacKunis, and
  Golubev]{pedroza2014robust}
Pedroza, N.; MacKunis, W.; Golubev, V.
\newblock Robust nonlinear regulation of limit cycle oscillations in uavs using
  synthetic jet actuators.
\newblock {\em Robotics} {\bf 2014}, {\em 3},~330--348.

\bibitem[Fischer \em{et~al.}(2014)Fischer, Hughes, Walters, Schwartz, and
  Dixon]{fischer2014nonlinear}
Fischer, N.; Hughes, D.; Walters, P.; Schwartz, E.M.; Dixon, W.E.
\newblock Nonlinear RISE-based control of an autonomous underwater vehicle.
\newblock {\em IEEE Trans. Robot.} {\bf 2014}, {\em 30},~845--852.

\bibitem[Economou \em{et~al.}(2000)Economou, Lee, Mavroidis, and
  Antoniadis]{economou2000robust}
Economou, D.; Lee, C.; Mavroidis, C.; Antoniadis, I.
\newblock Robust vibration suppression in flexible payloads carried by robot
  manipulators using digital filtering of joint trajectories.
\newblock  In Proceedings of the {International Symposium on Robotics and Automation}, Monterrey, N.L., Mexico, November 2000; pp. 244--249.  

\bibitem[Mamani \em{et~al.}(2012)Mamani, Becedas, and Feliu]{mamani2012sliding}
Mamani, G.; Becedas, J.; Feliu, V.
\newblock Sliding mode tracking control of a very lightweight single-link
  flexible robot robust to payload changes and motor friction.
\newblock {\em J. Vib. Control} {\bf 2012}, {\em
  18},~1141--1155.

\bibitem[Feliu \em{et~al.}(2013)Feliu, Castillo, Jaramillo, and
  Partida]{feliu2013robust}
Feliu, V.; Castillo, F.; Jaramillo, V.; Partida, G.
\newblock A Robust Controller for A 3-DOF Flexible Robot with a Time Variant
  Payload.
\newblock {\em Asian J. Control} {\bf 2013}, {\em 15},~971--987.

\bibitem[Gu \em{et~al.}(2005)Gu, Petkov, and Konstantinov]{gu2005robust}
Gu, D.W.; Petkov, P.; Konstantinov, M.M.
\newblock {\em Robust Control Design with MATLAB{\textregistered}}; Springer
  Science \& Business Media: Berlin/Heidelberg, Germany, 2005. 

\bibitem[From \em{et~al.}(2010)From, Schj{\o}lberg, Gravdahl, Pettersen, and
  Fossen]{from2010boundedness}
From, P.J.; Schj{\o}lberg, I.; Gravdahl, J.T.; Pettersen, K.Y.; Fossen, T.I.
\newblock On the boundedness and skew-symmetric properties of the inertia and
  Coriolis matrices for vehicle-manipulator systems.
\newblock {\em IFAC Proc. Vol.} {\bf 2010}, {\em 43},~193--198.

\bibitem[Dixon \em{et~al.}(2013)Dixon, Behal, Dawson, and
  Nagarkatti]{dixon2013nonlinear}
Dixon, W.E.; Behal, A.; Dawson, D.M.; Nagarkatti, S.P.
\newblock {\em Nonlinear Control of Engineering Systems: A Lyapunov-Based
  Approach}; Springer Science \& Business Media: Berlin/Heidelberg, Germany,  2013. 

\bibitem[Crane~III and Duffy(2008)]{crane2008kinematic}
Crane~III, C.D.; Duffy, J.
\newblock {\em Kinematic Analysis of Robot Manipulators}; Cambridge University
  Press: New York, USA, 2008.

\bibitem[Slotine \em{et~al.}(1991)Slotine, Li, et~al.]{slotine1991applied}
Slotine, J.J.E.; Li, W.; others.
\newblock {\em Applied Nonlinear Control}; Prentice Hall: Englewood Cliffs, NJ, USA,
  1991.

\bibitem[Khalil(2002)]{khalil2002nonlinear}
Khalil, H.
\newblock {\em Nonlinear Systems}; Prentice Hall: Englewood Cliffs, NJ, USA,  2002.

\bibitem[Desoer and Vidyasagar(1975)]{desoer1975feedback}
Desoer, C.A.; Vidyasagar, M.
\newblock {\em Feedback Systems: Input-Output Properties}; Academic Press: New York, USA
, 1975.

\bibitem[Lewis \em{et~al.}(2003)Lewis, Dawson, and Abdallah]{lewis2003robot}
Lewis, F.L.; Dawson, D.M.; Abdallah, C.T.
\newblock {\em Robot Manipulator Control: Theory and Practice}; Marcel Dekker: New York, USA,
  2004. 
\bibitem[Craig(1988)]{craig1988adaptive}
Craig, J.
\newblock {\em Adaptive Control of Mechanical Manipulators}; Addison-Wesley: Boston, MA, USA
 , 1988.

\bibitem[Kokotovic(1992)]{kokotovic1992joy}
Kokotovic, P.V.
\newblock The joy of feedback: Nonlinear and adaptive.
\newblock {\em IEEE Control Syst. Mag.} {\bf 1992}, {\em 12},~7--17.

\bibitem[Xian \em{et~al.}(2004)Xian, De~Queiroz, and
  Dawson]{xian2004continuous}
Xian, B.; De~Queiroz, M.S.; Dawson, D.M.
\newblock A Continuous Asymptotic Tracking Control Strategy for Uncertain Nonlinear Systems. \newblock {\em IEEE Trans. Autom. Control} {\bf 2004},  {\em
  49},~1206-1211. 

  
\end{thebibliography}
\end{document}